%Paper: hep-th/9212136
%From: ologhlin@physics.rutgers.edu (martin oloughlin)
%Date: Tue, 22 Dec 92 16:50:16 EST
%Date (revised): Tue, 22 Dec 92 17:33:12 EST
%Date (revised): Thu, 3 Jun 93 14:43:52 EDT

\input harvmac

%%%%%%%%%%%%%%%%%%%%%%%%%%%%%%%%%%%%%%%%%%%%%%%%%%%%%%%%%%%%%%%
%The following lines are needed to insert the accompanying figures in
%the paper. If you do not have epsf, then comment out the line
% ``\input epsf'', and print the figures separately. The figures are at
%the end of the tex file, with instructions for their extraction.
\input epsf
\ifx\epsfbox\UnDeFiNeD\message{(NO epsf.tex, FIGURES WILL BE IGNORED)}
\def\figin#1{\vskip2in}% blank space instead
\else\message{(FIGURES WILL BE INCLUDED)}\def\figin#1{#1}\fi
\def\ifig#1#2#3{\xdef#1{fig.~\the\figno}
\goodbreak\midinsert\figin{\centerline{#3}}%
\smallskip\centerline{\vbox{\baselineskip12pt
\advance\hsize by -1truein\noindent{\bf Fig.~\the\figno:} #2}}
\bigskip\endinsert\global\advance\figno by1}
%%%%%%%%%%%%%%%%%%%%%%%%%%%%%%%%%%%%%%%%%%%%%%%%%%%%%%%%%%%%%%%%

\Title{\vbox{\baselineskip12pt\hbox{hep-th/9212136}\hbox{RU-92-61}}}
{\vbox{\centerline{Nonsingular Lagrangians for Two
Dimensional Black Holes }
}}
\centerline{(Revised Version)}
\bigskip
\centerline{\it
T. Banks and M. O'Loughlin \footnote{*}
{\rm Supported in part by the Department of Energy under grant No.
DE-FG05-90ER40559  .} }
\smallskip
\centerline{Department of Physics and Astronomy}
\centerline{Rutgers University}
\centerline{Piscataway, NJ 08855-0849}
\noindent
\bigskip
%\vskip 2cm
\baselineskip 22pt
\noindent
We introduce a large class of modifications of the standard lagrangian
for two dimensional dilaton gravity, whose general solutions are
nonsingular black holes.  A subclass of these lagrangians have extremal
solutions which are nonsingular analogues of the
extremal Reissner-Nordstrom spacetime.
It is possible that quantum deformations of
these extremal solutions are the
endpoint of Hawking evaporation when the models are coupled to matter,
and that the resulting evolution may be studied entirely within the
framework of the semiclassical approximation.
Numerical work to verify this conjecture is in progress.  We point out
however that the solutions with non-negative mass
always contain Cauchy horizons,
and may be sensitive to small perturbations.

\Date{December 1992}
%\draftmode

\baselineskip 26pt
\newsec{Introduction}

\lref\GMGHS{G. Gibbons and K. Maeda, {\it Nucl.
Phys.}{\bf B298}(1988), 741; D. Garfinkle, G. Horowitz, and A.
Strominger, {\it Phys. Rev.}{\bf D43},(1991),3140; Erratum: {\it Phys. Rev.}
{\bf D45} (1992), 3888.}
\lref\bddo{T. Banks, A. Dabholkar, M.R. Douglas, and M. O'Loughlin, ``Are
horned particles the climax of Hawking evaporation?'' {\it Phys. Rev.}
{\bf D45}(1992),3607.}
\lref\corn{S.B. Giddings and A. Strominger, ``Dynamics of
extremal black holes'', {\it Phys. Rev.} {\bf D46} (1992) 627;
M.Alford,A.Strominger, ``S Wave Scattering of Charged
Fermions by a Magnetic Black Hole'', {\it Phys.Rev.Lett.}{\bf
69},(1992),563.}
\lref\bol{T. Banks and M. O'Loughlin,
``Classical and quantum production of cornucopions at energies below
$10^{18}$ GeV'', Rutgers preprint RU-92-14 (1992).}
\lref\gidstrom{S. Giddings and A. Strominger, ``Dynamics of extremal
black holes'',{\it Phys. Rev.}{\bf D46} (1992), 627.}
\lref\bghs{B. Birnir, S. Giddings,
J. Harvey, A. Strominger, ``Quantum Black Holes''
{\it Phys. Rev. }{\bf D46}(1992), 638.}
\lref\hawk{S. Hawking, ``Evaporation of two-dimensional black holes'',
{\it Phys. Rev. Lett.}{\bf 69}(1992), 406.}
\lref\sthandb{L. Susskind, L. Thorlacius,
``Hawking Radiation and Back-Reaction'', {\it Nucl. Phys. }{\bf
B382}(1992), 123.}
\lref\hastew{S.W. Hawking, J.M. Stewart, ``Naked and Thunderbolt
singularities in Black Hole Evaporation'' Cambridge preprint,
PRINT-92-0362, hep-th/9207105 (1992).}
\lref\dlowe{D.A. Lowe, ``Semiclassical approach to Black Hole
Evaporation,''Princeton preprint, PUPT-1340, hep-th/9209008 (1992).}
\lref\rst{J. Russo, L. Susskind, and L. Thorlacius, ``
Black hole evaporation in 1+1 dimensions'' {\it Phys. Lett.}{\bf B292}(1992),
13.}

Recently, there has been significant progress in unravelling the mystery
which enshrouds the endpoint of Hawking evaporation of black
holes \ref\everyone{C.G. Callan, S.B. Giddings, J.A. Harvey, and A.
Strominger, ``Evanescent black holes,''  {\it Phys.
Rev.}{\bf D45} (1992) R1005; J. Harvey and A. Strominger, ``Quantum
Aspects of Black Holes'', Enrico Fermi Institute preprint EFI-92-41,
hep-th/9209055, and references cited
therein.}.  In particular, we now believe that the simple
arguments that appeared to rule out stable remnants as a plausible
endpoint for black hole evaporation, are wrong.  The context in which
these ideas have been developed was that of extremal magnetically
charged black holes in the version of gravity (dilaton gravity) which
appears in the low energy limit of string theory \refs{\GMGHS ,\bddo,
\corn ,\bol}.  It has long been
argued that extremal charged black holes might be the natural final
state for a black hole that manages to retain its charge in the process
of Hawking evaporation.  In the case of dilaton gravity, the geometry of
the extremal magnetic black hole (shown in fig.1) is completely static,
horizon free and has no singularities at finite points of space. It has
the form  of an infinite funnel or
horn, attached to an asymptotically flat space.
The only singularity of the solution is
the divergence of the effective coupling an infinite distance down the
horn.

\ifig\fone{The spatial geometry of an extremal dilatonic magnetic black hole.
The cross sections of the throat are two spheres.}
{\epsfysize=3.0in\epsfbox{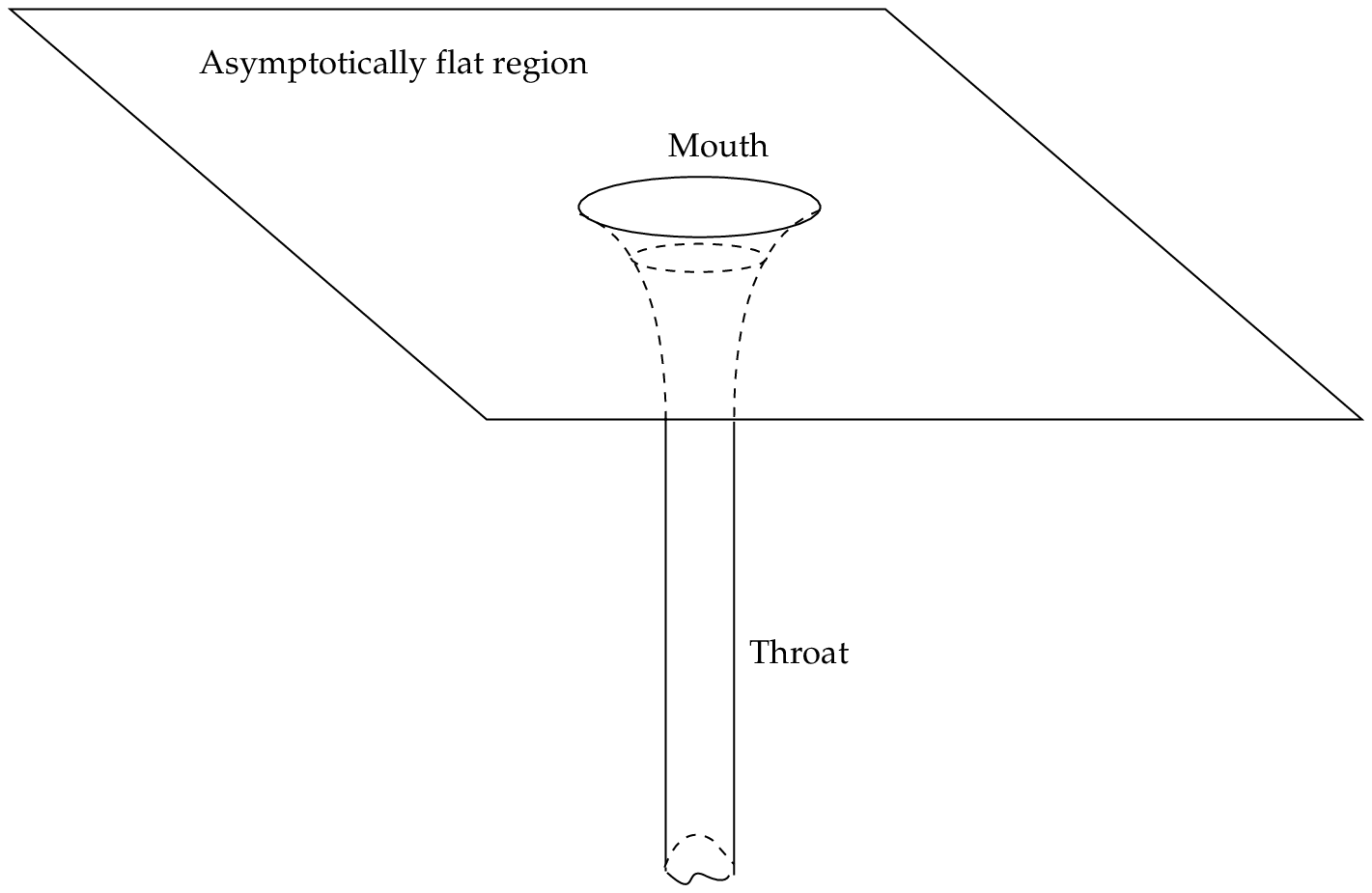}}

Although an external observer with sufficiently coarse resolving power
will see such a black hole as a pointlike object, it is not an
elementary particle.  In a previous publication \bddo we have named such
objects horned particles or cornucopions.
 The infinite volume of the horn is a repository
for an infinite number of states of quantum fields propagating on the
background geometry. It has been argued that states which differ only by
excitations localized far down the horn, will be essentially degenerate
in ADM energy.  The extremal hole is thus a candidate for the kind of
infinitely degenerate remnant which might resolve the information loss
paradox of Hawking evaporation.  Its large size also allows it to evade
the apparent phenomenological problems of infinite production cross
sections, and infinite contributions to virtual loops
that usually plague the idea of black hole relics \ref\BOS{T. Banks, A.
Strominger and M. O'Loughlin, ``Black Hole
Remnants and the Information Puzzle'', Rutgers Preprint RU-92-40,
hep-th/9211030.}.

\lref\lindeguth{E. Farhi,A. Guth, ``Is it possible to create a
universe in the laboratory by quantum tunneling?'' {\it Phys.
Lett.} {\bf B183},(1987),149; A. Linde, ``Stochastic approach to tunneling
and baby universe formation'', {\it Nucl. Phys. }{\bf B372} (1992) 421.}
\lref\frolov{V.P.Frolov,M.A.Markov and
V.F.Mukhanov, ``Black Holes as possible Sources of Closed and
semi-closed universes'', {\it Phys. Rev.}{\bf D41} (1990),
383.}
\lref\morgan{D. Morgan, ``Black Holes in Cutoff
gravity''{\it Phys. Rev.} {\bf D43},(1991),3144.}
It is important to emphasize that the conceptual picture that has been
built up for magnetically charged black holes is, in general terms,
equally applicable to neutral holes.  In the charged case we can find a
completely classical picture of the remnant\foot{This is somewhat
illusory.  The weak coupling approximation breaks down deep in the
hole.}, but one can conjecture that a similar picture (with a bit of
quantum fuzz near the region of the classical singularity) might be
applicable to neutral holes as well.
The idea of geometries in which the Schwarzchild singularity
is replaced by an
expanding internal universe \refs{\lindeguth ,\frolov} is an old one.
Linde has argued that black holes with
internal universes appear naturally in the chaotic inflation scenario,
and Farhi and Guth proposed creating one in the
laboratory \lindeguth .  While such a scenario is not consistent
with the classical Einstein field equations when matter satisfies the
dominant energy condition, it could arise due to quantum
effects \refs{\frolov ,\morgan}.  In many of these discussions,
the universe on
the other side of the black hole throat is taken to be a DeSitter space.
Strominger\ref\ghost{A. Strominger, ``Fadeev-Popov ghosts and 1+1
dimensional black hole evaporation,'' {\it Phys. Rev. D}, to appear,
hep-th /9205028.}
 has argued that nonsingular black hole geometries with an
internal DeSitter structure might appear in the solutions
of a particular class of semiclassical equations describing two
dimensional black hole evaporation.

The space times envisaged in these
scenarios differ quantitatively but not qualitatively from the
cornucopion.  The geometry beyond the mouth of the
latter can be viewed \bol as the limit of an
anisotropically and inhomogeneously expanding four dimensional universe
with the shape of an expanding cigar cross a two sphere of constant
radius.  In the spacetimes of the previous paragraph, the internal
geometry far from the throat, is isotropic and uniformly expanding.
The common element in all these scenarios, is that black hole
formation results in the production of a new asymptotic region of space,
separated from the old one by a neck of small size.  While such a
spacetime can be foliated by spacelike surfaces, and the evolution of
quantum fields between these surfaces is unitary, the S-matrix of the
initial asymptotic region is not.  An observer in this region loses
information into what are for her the ``internal states'' of a pointlike
remnant.  She can construct a sensible quantum mechanics of these
remnants without examining their internal structure, as long as she does
not study processes in which they are created or annihilated.  The latter
require her to understand the internal structure of the remnant \BOS .

While the general picture provided by these model spacetimes is
satisfying, a crucial part of the puzzle remains unsolved.  To date, no
one has demonstrated the collapse of a nonextremal black hole to one of
the hypothetical remnants into which it is supposed to evolve.  Clearly
such a demonstration will require us to understand physics at a level
beyond the classical Einstein theory of gravity, but it is not clear
what extension of Einstein's theory is necessary.  The essence of the
problem is the black hole singularity, and this is widely viewed as a
problem having to do with short distance physics.  String theory is the
most plausible candidate for a short distance extension of Einstein's
theory and there have been repeated suggestions that the existence of a
fundamental length in classical string theory might eliminate black hole
singularities.\foot{Witten \ref\wit{E. Witten, ``On string theory and
black holes,''
{\it Phys. Rev. }{\bf D44}(1991),314.} has argued that his exact classical
black hole solution of two dimensional string theory shows that this
conjecture is false.  We feel that too little is known about the actual
properties of this solution to justify this conclusion.}  On the other
hand, it has often been argued that the high curvature region near the
singularity of a collapsing black hole will be a region where quantum
fluctuations are very important.  One might hope for example that
quantum fluctuations in some scalar field near the singularity, create a
locally large value of the cosmological constant, setting off chaotic
inflation inside the black hole, and producing one of the DeSitter
remnants described above.  While these ideas are alluring, they do not
lend themselves to systematic mathematical investigation.

 The main
advantage of studying extremal dilaton black holes, rather than neutral
ones, is that one can plausibly argue \ref\cghs{C.G. Callan, S.B.
Giddings, J.A. Harvey, and A. Strominger, ``Evanescent black holes,''
{\it  Phys. Rev.}{\bf D45} (1992), R1005.} that the collapse and
evaporation of near extremal dilaton
black holes can be entirely described by an
effective two dimensional field theory.  In this effective theory, the
black hole singularity can (by an appropriate, and string theoretically
natural, choice for the conformal frame of the metric) be attributed
entirely to the blow up of the coupling constant.  One is led to hope
that a fully quantum mechanical solution of the low energy effective field
theory may be all that is necessary to an understanding of the
singularity\foot{This is the case in $1 + 1$ dimensional string theory.
Although we still lack a proper nonperturbative definition
of this theory, it is clear that the singularity of the classical
solution with vanishing tachyon condensate is eliminated by quantum
mechanics.}.  All extant attempts to solve this problem have relied on
the quantum fluctuations of a set of massless degrees of freedom, and
treated the metric and dilaton as classical mean fields.  This approach
has led to the first explicit description of Hawking radiation with back
reaction included, but the semiclassical metric becomes singular (now at
a finite value of the coupling), and the mean field expansion breaks
down\everyone .

There are two kinds of corrections to the mean field description which
might become important in the strong coupling regime.  The first are quantum
fluctuations of the metric and dilaton,
and within the CGHS model these are the only corrections to mean field
theory.  It is
obviously a problem of great conceptual interest to learn how to treat these
fluctuations correctly, but it is also a very difficult problem.
If we view the CGHS model as the low energy effective theory of a more
complete Lagrangian, then there are other quantum corrections that might
be equally important.  These corrections, first discussed in \bddo
come from integrating out quantum
fluctuations of the heavy fields in the full Lagrangian.  In effective
field theory language they correspond to relevant and marginal operators
in the low energy theory which have different scaling behavior than the
CGHS Lagrangian when the dilaton is shifted by a constant.  An $l$
loop contribution will scale like $e^{2 l\phi}$.
If we treat the graviton and dilaton fields classically, there are an
infinite number of relevant and marginal operators that can be added to
the Lagrangian, the most general renormalizable Lagrangian having the
form \ref\BOone{T.Banks,M.O'Loughlin, ``Two-dimensional quantum gravity in
Minkowski space'', {\it Nucl. Phys.}{\bf B362},(1991), 649; S.D.
Odintsov and I.L. Shapiro, ``One-loop renormalization of two-dimensional
induced quantum gravity'', {\it Phys.Lett.}{\bf B263}(1991),183.}
\eqn\lagone{{\cal L} = \sqrt{-g}(D(\phi ) R + G(\phi )(\nabla\phi )^2 +
H(\phi )).}
with
\eqn\bc{D(\phi )\rightarrow {G(\phi )\over 4} \rightarrow H(\phi
)\rightarrow e^{-2\phi} }
as $\phi\rightarrow -\infty$.
In conformal gauge, with conformal factor $e^{2\rho}$, this is a
nonlinear model with a two dimensional Minkowski signature target space.
The $\rho$ direction is lightlike \ref\BL{T. Banks and J. Lykken,
``String Theory and two-dimensional
quantum gravity'', {\it Nucl. Phys.}{\bf B331},(1990),173;
S.R. Das, S. Naik, S.R. Wadia, ``Quantization of the Liouville Mode and
String Theory'', {\it Mod.Phys.Lett.}{\bf A4},(1989),1033.}.  The
renormalization group
equations for such a model are hyperbolic, and their solutions are
determined by the initial data on a surface of constant $\rho$ given by
the three functions in \lagone .

It is a formidable task to find
the functions $D$, $G$, and $H$ that are obtained by integrating out heavy
degrees of freedom in a realistic model.  In this paper we will simply
explore the possibilities, with a view to answering the following
question: is it possible for the quantum corrections to the effective
action to smooth out the strong coupling singularity of the CGHS
lagrangian? If this is the case, we might be able to solve the black
hole evaporation problem without dealing with the intricate
interpretational problems of the quantum theory of gravity.  The mean
field approach pioneered by CGHS would be sufficient.  Of course,
finding a particular lagrangian which avoids singularities is not very
satisfactory.
 In view of our
ignorance of the form of the realistic corrections to the effective
lagrangian, it would be well to find that the presence or absence of
singularities was a generic property of large classes of lagrangians of
the form \lagone .

We will show in section II that a very large subclass of the
lagrangians of \lagone have solutions which asymptote to the the CGHS
black holes for weak coupling, but are completely nonsingular space
times which are geodesically complete as the coupling goes
to infinity.  Thus, for these lagrangians, the infinite coupling
singularity is located at an infinite geodesic distance for any value of
the black hole mass.  The rather generic occurence of nonsingular black
holes with internal DeSitter asymptotics is interesting, but by itself
does not prove that a semiclassical description of Hawking evaporation
is possible.  We must face
another important issue.  In four dimensional Einstein gravity and the
CGHS model, the nonsingular ``vacuum'' solution is the boundary between
solutions which have naked singularities and those whose singularity is
hidden from the asymptotic region by a horizon.  The
positive energy theorem holds only for solutions which do not have naked
singularities.  For many of our nonsingular lagrangians, there is no
qualitative difference between solutions with positive and negative ADM
mass.  Thus one might expect that once our models are coupled to
dynamical matter fields, so that black holes can radiate, the radiation
will go on forever, down to infinitely negative energy\foot{The
importance of this potential problem was brought to our attention by A.
Strominger.}.  Precisely such
a disaster occurs in exactly soluble semiclassical models of black hole
evaporation which exploit Strominger's \ghost  mechanism for
avoiding singularities \ref\negen{A. Bilal and C. Callan, ``Liouville
Models of Black Hole
evaporation'', {\it Nucl.Phys.}{\bf B394},(1993),73;
S.P. de Alwis, ``Black Hole Physics from Liouville
Theory.''{\it Phys.Lett.}{\bf B300},(1993),330;
S.P. de Alwis, ``Quantization of a theory of 2d dilaton
gravity'', {\it Phys.Lett.}{\bf
B289},(1992),278; S.B. Giddings, A. Strominger, ``Quantum Theories of
Dilaton Gravity'' {\it Phys.Rev.}{\bf D47},(1993),2454.}.

It turns out that one can obtain nonsingular
lagrangians which have a dichotomy between ``positive'' mass and
``negative'' mass solutions.  We put
the terms positive and negative
 in quotes because we measure the ADM mass of all of these solutions
 relative to a certain extremal solution of the equations rather than
to the linear dilaton vacuum of the
CGHS model.  We believe that this is the correct procedure.
The linear dilaton
vacuum is not a solution of any of the modified lagrangians that we have
studied.  The proper reference point for ADM mass is the stable solution
of the equations which we expect to be the endpoint of Hawking
evaporation.
  We will explain the
detailed geometries of these solutions in a more leisurely manner in
section III.

In section IV we briefly discuss the semiclassical equations which arise
by coupling the models of III to massless CGHS ${\bf f}$ fields in the leading
order of a certain large $N$ expansion.  This discussion is preliminary,
as we have not yet solved the equations.  We conclude by outlining the
further steps which must be taken to carry out the analysis which we
have begun in this paper.

\newsec{Nonsingular Lagrangians and Negative Energy}

In \BOone we wrote down the most general renormalizable Lagrangian for
the two dimensional metric and a single scalar field.  For a general
choice of field variables it takes the form \lagone .  We will require
that the coupling functions satisfy the CGHS boundary condition \bc .
The metric in these Lagrangians is by definition the stringy metric,
which is distinguished by its simple coupling to propagating
strings\foot{There is in fact an ambiguity in what we mean by the
stringy metric arising from renormalization scheme ambiguities in world
sheet $\sigma $ models.  This problem arises in higher orders in the
string tension expansion and should not effect the low energy
considerations of this paper.}.  For purposes of solving the equations of
motion, it is convenient to perform a Brans-Dicke transformation to
eliminate the $G$ term in the Lagrangian.  This can be done in a
nonsingular way whenever ${d D \over d\phi} \neq 0$ over the entire range
of $\phi$.  For the CGHS Lagrangian, this criterion is satisfied
for any finite value of the string coupling, $e^{2\phi}$, but fails when
the coupling goes to infinity.  As we will see in a moment,
it is no accident that this is also the
locus of singularities of the classical solutions of this Lagrangian.

The appropriate Brans Dicke transformation is
\eqn\bdtrans{g_{\mu\nu}^{CGHS} = e^{2 S(\phi )} g_{\mu\nu}}
where
\eqn\bdcond{4{d S \over d\phi}  {d D \over d\phi} = - G(\phi )}
If $W \equiv H e^{2S}$, then the Lagrangian of the transformed fields is
\eqn\bdlag{\sqrt{-g}(D R + W)}
It is amusing to note that the field equation for $\phi$ which
 follows from
this Lagrangian involves no derivatives.  Thus this lagrangian is
equivalent to a higher derivative Lagrangian involving only the
gravitational field
\eqn\hdlag{{\cal L}_{HD} =\sqrt{-g} {\cal F} (R)}
Note however that the function ${\cal F}$ results from inverting the
functional relation between $\phi$ and $R$ given by the field equations,
so that it might be multivalued.

To find the general solution of the field equations, we recall our
general result \BOone that any such solution has a Killing vector, and
that the dilaton is constant along the Killing flows.  If
$t$ is a coordinate along the homogeneous direction, and $\sigma$ a
coordinate orthogonal to it, we can choose $\sigma$ so that the metric is
conformally flat, with conformal factor $e^{2 \hat\rho (\sigma )}$
and the dilaton is
a function only of $\sigma$.  In this coordinate system, the constraint
equations reduce to a single equation (dots refer to $\sigma$ derivatives):
\eqn\constr{\ddot D = 2{\dot{\hat\rho}}\dot D}
while the variational equation for the conformal factor is:
\eqn\vareqn{{\ddot D} =  e^{2\hat\rho} W}
The first equation is solved by
\eqn\statsola{{\dot D} = 2\beta  e^{2\hat\rho},  }
where $\beta$ is a constant.  Plugging this back into the second
equation we get

\eqn\hofsig{ h(\phi ) \equiv {d\phi \over d\sigma}(\phi ) = {1 \over {2 \beta
D^\prime}} \int d\phi \quad W(\phi ) D^\prime (\phi )}

Here, primes denote derivatives with respect to $\phi$.
Thus, the solution of $\phi$ in terms of $\sigma$ is reduced to
quadratures.
Note from \hofsig  that generically $\dot\phi = 0$
at a number of values of $\sigma$ equal to the number of zeroes of
$W(\phi)D^\prime (\phi)$, plus one.

The two qualitative features of these solutions
that we would like to discuss at this point
are, the behaviour of the solutions at the horizon, and the behaviour in
the strong coupling region.
In the coordinates in which
we are working, an apparent horizon is a point where
($\rho$ is the Liouville field in the CGHS frame)
$e^{2 \rho} = 0 $ and $\dot\phi = 0$.  The fact that these two
conditions coincide for finite values of $\phi$ follows from the equation
\eqn\cghsmet{e^{2\rho} = e^{2\hat\rho + 2\phi} =  {D^\prime
(\phi )\dot\phi e^{2\phi} \over {2\beta}}}
and our fundamental assumption that the derivative of $D$ vanishes
nowhere.  When $h$ has a linear zero, the behavior of the solution is
exactly the same as that of the standard dilaton gravity black hole.
The contours of constant dilaton field change from being spacelike to
timelike or vice versa, but the dilaton is monotonic across the horizon.
It is also of course monotonic in regions where $h \neq 0$.  Thus, the
sort of ``bounce'' solutions that appeared in the static quantum
equations of \refs{\bghs ,\sthandb ,\hawk} are not obtained in any of
the models we are studying.

To find singularities, we express the curvature of the CGHS
metric in terms of our solutions:

\eqn\scalarcurvR{\eqalign{
R_{CGHS} &= 8 e^{-2\rho}\del_+ \del_- \rho = -4 \beta
{e^{-2\phi} \over {h D^\prime}} \ddot\rho
\cr &= -{2 \beta e^{-2\phi} \over D^\prime}\bigl({h^\prime D^{\prime\prime}
\over D^\prime} + 2 h^\prime + h^{\prime\prime} +
{D^{\prime\prime\prime} h \over D^\prime} - {(D^{\prime\prime})^2 h
\over (D^\prime)^2}\bigr)
}}

This is finite at any finite value of $\phi $ as long as $D$ and $W$ are
smooth, and $D^\prime$ does not vanish.  In the weak couping region
$\phi\rightarrow -\infty$, $R$ goes to zero because of our requirement \bc
that the lagrangian approach that of CGHS.  Let us assume that in the
strong coupling region, $D \sim e^{n\phi}$ and $W \sim e^{m\phi}$.  Then
$h \sim e^{m\phi}$, and barring accidental cancellations,
$R_{CGHS} \sim e^{(m - n -2 ) \phi}$, so that, if $n \geq m - 2$, it remains
finite when the coupling goes to infinity.

We can also examine the geodesic distance to the infinite coupling
region:
\eqn\geodist{\int e^\rho d\sigma = \int \sqrt{D^\prime h} e^\phi d\sigma
=\int \sqrt{D^\prime \over h} e^\phi d\phi \sim e^{{(n - m + 2) \over 2}
\phi}}
This is infinite whenever $n \geq m - 2$. Note by the way that the
distance to any apparent horizon from a point where $\phi$ takes on a
general finite value, is generically finite. Only if $h(\phi)$ has a
double zero will it be infinite. This observation will be important in
the sequel.

We have thus exhibited a large class of models whose general black hole
solution is no more singular than the linear dilaton vacuum of the CGHS
action.  Indeed, they are less singular.  Although the ``string
coupling'', $e^{2\phi}$, becomes infinitely strong asymptotically in our
solutions, this no longer signals a region of large quantum fluctuation.
Quantum fluctuations in the graviton and dilaton fields are controlled
by the value of $1\over D^\prime (\phi )$, which is bounded even in the
``strong coupling'' region.

 To conclude this section, we record a simple set of Lagrangians
which generate nonsingular black holes with asymptotically DeSitter
interiors.  We take $D = e^{-2\phi} - {\gamma^2 \over n}e^{2n\phi}$ and $W =
 - 2 \lambda^2 D^{\prime} (\phi ) e^{2\phi}$.  This gives (with $\beta =
 \lambda$, obtained by choosing asymptotically linear dilaton co-ordinates),
\eqn\hone{h(\phi) = {2\lambda \over{(1 + \gamma^2 e^{2(n+1)\phi})}}\bigl(-
\half + {M \over {2\lambda}} e^{2\phi} + {\gamma^2 \over n}
e^{2(n+1)\phi} + {\gamma^4 \over {(4n + 2)}}e^{4(n + 1)\phi}\bigr ).}
Since this has only a single real zero, the metric has one horizon.  The
metric and dilaton are given by,
\eqn\sigofphi{\sigma (\phi)={1\over 2\lambda} \int d\phi
{(1+\gamma^2 e ^{2(n +
1)\phi}) \over (-\half + {M \over 2\lambda}e^{2\phi}  + {\gamma^2 \over
n} e^{2(n+1)\phi} + {\gamma^4 \over (4 n +2)}e^{4(n+1)\phi})}}
Asymptotically, and in the vicinity of the horizon we find,
\eqn\desitone{\sigma (\phi) = \left\{\matrix{-{\phi\over
\lambda} -  {M\over 2\lambda^2}e^{2\phi} + \dots,&\phi \rightarrow
-\infty\cr &\cr
{\alpha \over 2\lambda} log(\phi - \phi_0),&\phi
\rightarrow \phi_0 \cr &\cr -{(2n + 1) \over 2\lambda \gamma^2(n+1)}
e^{-2(n + 1)\phi},&\phi \rightarrow \infty}\right.}
Inverting to obtain metric and dilaton,
\eqn\desittwo{\phi (\sigma) =\left\{\matrix{-\lambda \sigma - {M \over
2\lambda}e^{-2\lambda \sigma} + \dots ,&\phi \rightarrow -\infty\cr
&\cr \phi_0 - e^{2{\lambda\over \alpha} \sigma } + \dots, &\phi
\rightarrow \phi_0 \cr &\cr -{1\over 2(n+1)}log(\sigma_\infty -
\sigma),&\phi \rightarrow \infty}\right.}
\smallskip
\eqn\desitthree{e^{2\rho} = \left\{\matrix{1 -
{M\over \lambda}e^{-2\lambda\sigma} + \dots,&\phi \rightarrow -\infty\cr
&\cr e^{{2\lambda\over\alpha}\sigma},& \phi \rightarrow \phi_0\cr
&\cr {const \over (\sigma - \sigma_\infty)^2},&\phi \rightarrow
\infty}\right.}
They exhibit the advertised asymptotic DeSitter behavior in the strong
coupling region.

\newsec{Models With Semiclassically Stable Solutions?}

Although we have exhibited a general class of effective Lagrangians with
nonsingular black hole solutions, we do not think that the results of
the previous section constitute a demonstration that we can construct
sensible semiclassical models of black hole evaporation.  In a sense, we
have done our work too well.  The models that we have constructed have
nonsingular solutions, for all values of the ADM mass, {\it including
negative ones}.  One suspects that when the black holes of these models
are coupled to matter, they will suffer the fate of the models
of \negen .  Hawking evaporation will proceed forever, leaving
behind nonsingular black holes of ever larger negative ADM mass.

To see
why we anticipate this disaster, consider the class of models, which are
nonsingular according to the criteria of the previous section, and in
which $W(\phi )$ has no zeros.  $h(\phi )$ then has a single zero, which
is a finite geodesic distance from points with finite $\sigma$
coordinates.  This is true for all finite positive and negative values
of the ADM mass.  The asymptotically timelike
Killing vector of the solution is null at this horizon, and becomes
spacelike on the other side of it.  It
is easily seen in the semiclassical approximation
that matter propagating in any of these geometries will Hawking radiate.
This makes it seem implausible that the equations
with backreaction included will have solutions which become
asymptotically static.  There is no apparent reason for the radiation to
turn off.

\ifig\ftwo{Extended Penrose diagram of the classical metric for the case
of two zeroes in $h$. Regions $I$ and $III$ are static, but region $II$, is
time-dependent. The arrows indicate the direction of the Killing vector
used to obtain these solutions. The line AB indicates a Cauchy horizon
for region $I_a$. The double line indicates the region that is
asymptotically linear dilaton vacuum.}
{\epsfysize=3.0in\epsfbox{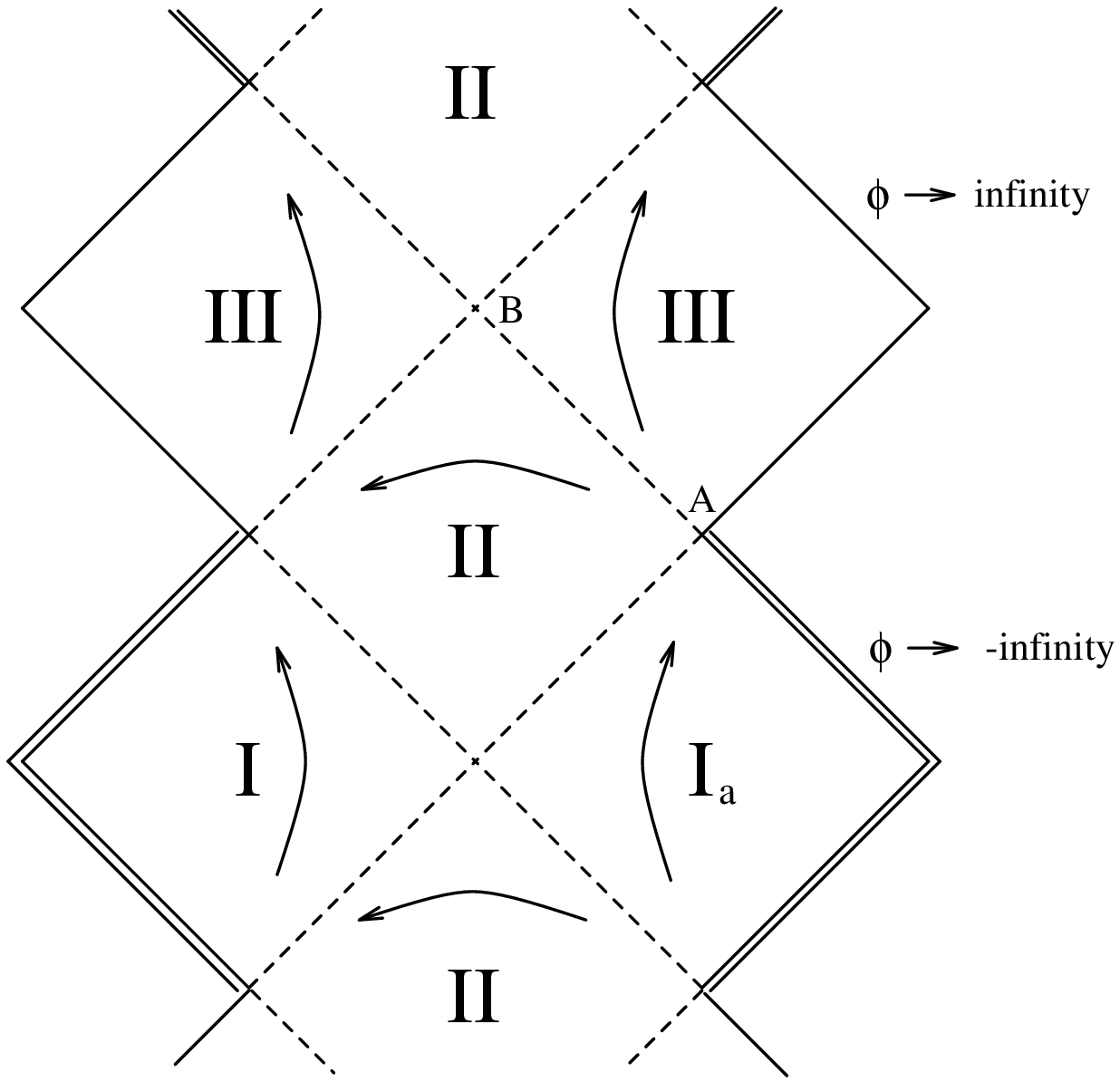}}

While this is not a proof that these models suffer the fate of those
studied in \negen , we were sufficiently convinced of this to search
for models in which the horizon moved off to infinity at some finite value of
ADM mass.  Such models can be constructed by choosing $W(\phi )$ to
change sign once in the physical range of $\phi$.  For the generic
solution, $h(\phi )$ then has two zeroes, and there is a particular
value of the ADM mass for which these zeroes coincide.  If we impose
restrictions on the strong coupling behavior of $D$ and $W$ which
guarantee the absence of singularities, the generic
geometry has two half infinite static regions, connected through a pair
of horizons, to a homogeneous expanding universe.  This
is not the complete geometry however, since there are geodesics which
enter into the expanding region and never come back.  The second horizon
is a Cauchy horizon for the region connected to weak coupling.  If we
follow the usual procedure of analytically continuing through this
Cauchy horizon, we obtain
the full Penrose
diagram shown in \ftwo .  It has the periodic structure
familiar from the Kerr and Reissner-Nordstrom geometries, but is everywhere
nonsingular.   Parts of this spacetime can be foliated by spacelike
surfaces, but it contains Cauchy horizons.  This may mean that most of
the structure is unstable to small perturbations.

For the extremal value of the ADM mass, the two horizons coincide, and
the extended space-time now has zero Hawking
temperature. The causal structure is identical to that of the $M^2 =
Q^2$ extremal Reissner-Nordstrom solution, except that the timelike
singularity has been eliminated.
To see this explicitly, let us investigate the behavior of
the metric near a general zero of $h$:

\eqn\heqzero{h(\phi) = - \alpha^2 (\phi_0 - \phi)^k}

Integrating to get $\sigma(\phi)$, we obtain,
\eqn\sigofphi{\eqalign{\sigma &= \sigma_0 - {(\phi_0 - \phi)^{-(k - 1)}
\over {\alpha^2 (k-1)}}, k \neq 1 \cr &= \sigma_0 + {log(\phi_0 -
\phi)\over \alpha^2}, k = 1}}
Using \statsola , we can write for the metric ($k\neq 1$),
\eqn\metofphi{\eqalign{e^{2\rho} &= {h D^\prime e^{2\phi} \over
{2\lambda}}\cr &\approx
(\sigma - \sigma_0)^{-k\over{(k-1)}}, \sigma \to \sigma_0, k<1,\cr
&\approx \sigma^{-k\over{(k-1)}},\sigma\to-\infty,k>1.}}

We can see then that, for $k\geq 2$, $e^\rho$ cannot be integrated
through $\phi = \phi_0$ and so the distance
to $\phi = \phi_0$ is infinite, while for $k<2$ the distance to this
point is finite. (For $k=1$, $h(\phi) = e^{\alpha^2 \sigma}$, $e^{2\rho}
\approx e^{\alpha^2\sigma}$, and $\phi \to\phi_0$ as $\sigma\to
-\infty$). The curvature is given by $e^{-2\rho}\partial_\sigma^2\rho$,
and using the above expressions is easily seen to be infinite for
$1<k<2$ and $k<1$, finite for $k=2$ and zero for $k=1$. Note in particular that
$k=1$ near $\phi = \phi_0$ corresponds to the behaviour at the horizon
of the standard dilaton-gravity black hole.

If our effective action really comes from integrating out massive
fields, then we expect it to be an analytic function of the
dilaton\foot{Unless some heavy degrees of freedom became massless at a
particular value of $\phi$.  This phenomenon would complicate our
analysis, and we assume that it does not occur.}.  Thus $k = 1,2$ are
the only sensible nonsingular choices.  In fact, we expect simple zeroes
($k = 1$) of $h$ to be generic.  If $h$ has a single such zero, then we
have the sort of model analyzed in the previous section, which has the
potentially disastrous problem of runaway Hawking radiation.  If $W$ has
one or more simple zeroes, then $h$ has
multiple zeroes. By tuning the one parameter at our disposal, the
ADM mass, we expect to find a unique value of ADM mass at which the two zeroes
nearest to the weak coupling region coincide.  The existence of special
solutions for which the horizons coincide (as in the extremal
Reissner-Nordstrom black hole) is thus generic, as long as $W(\phi )$ has
one or more simple zeroes.
{\it The extremal geometry
is thus a nonsingular space-time with zero Hawking temperature.}
Quantum fields placed in such a gravitational field will not Hawking
radiate.

This observation  by itself does not guarantee that the
extremal geometry will furnish a satisfactory semiclassical endpoint for
Hawking evaporation.  Unlike the linear dilaton solution of the
classical CGHS model, the extremal solution has nonzero curvature and
will not be a solution of the one loop corrected mean field equations
which describe Hawking evaporation.  We must enquire whether there is a
static solution to these equations with properties similar to the
extremal geometry, and whether perturbations of this solution by
infalling matter eventually relax back to it.  We will set up the
equations which must be solved in the next section.  We have not yet
solved them.

Before concluding this section, we will work out the details of a
particular form of the Lagrangian which has solutions of the type we
have discussed.  We take
\eqn\dtwo{D = e^{-2\phi} - \ha \gamma^2 e^{4\phi}}
and
\eqn\wtwo{W = 4\lambda^2 - \mu e^{4\phi}}
Then
\eqn\htwo{ h = {2\lambda \over(1 + \gamma^2 e^{6\phi})}(-\half +
{M\over 2\lambda} e^{2\phi} -{\mu\over 8\lambda^2} e^{4\phi} + {\gamma^2
\over 4}e^{6\phi} -
 {\mu\gamma^2 \over 32 \lambda^2} e^{10 \phi})}

\ifig\fthree{A plot of $h(e^{2\phi})$ showing the crossover from two to no
zeroes. The conjectured extremal zero temperature solution lies at the
crossover point.}
{\epsfysize=2.0in\epsfbox{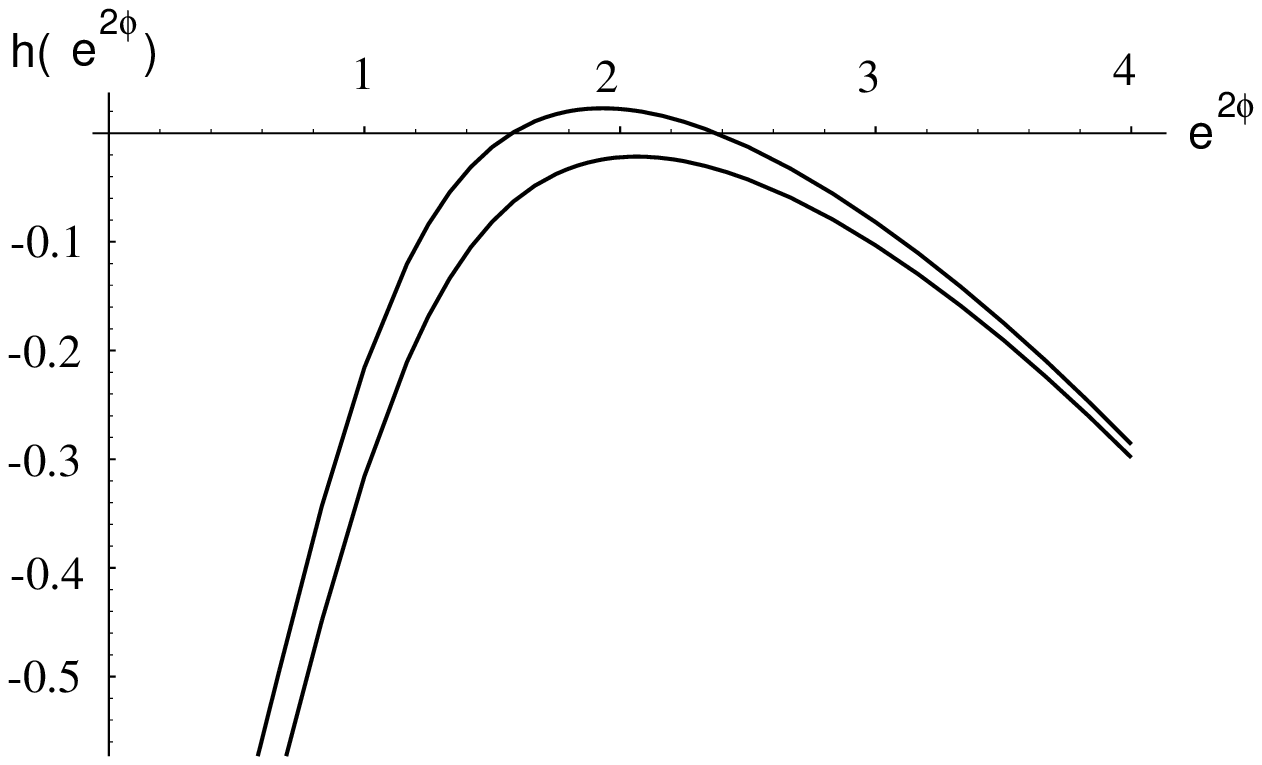}}

where $M$ is the ADM mass.  It is easy to see that the number of zeroes of
$h$ jumps from zero to two at a critical value of $M$, (see \fthree \quad
for an illustration for $\lambda
=\gamma =\mu=1$).  This is the extremal geometry of
this model.  The explicit asymptotic formulas for the metric and dilaton
at $\phi\rightarrow\pm\infty$, (when $h(\phi)$ has a zero the behaviour at
the zero will be typically of type $k=1$ discussed above, and in the
extremal case, $k=2$), are,
\eqn\sigofphi{\sigma (\phi)={1\over 2\lambda} \int d\phi
{(1+\gamma^2 e ^{6\phi}) \over (-\half + {M \over 2\lambda}e^{2\phi}  -
{\mu \over 8\lambda^2} e^{4\phi} + {\gamma^2 \over 8}e^{6\phi} -
{\mu\gamma^2 \over 64\lambda^2} e^{10\phi})}}
\noindent Giving,
\eqn\extrone{\sigma (\phi) = \left\{\matrix{-{\phi\over
\lambda} - {M\over 2\lambda}e^{2\phi} + \dots,&\phi \rightarrow
-\infty\cr &\cr \sigma_\infty +
{8\lambda\over\mu}e^{-4\phi}&\phi \rightarrow \infty}\right.}
and inverting this expression to obtain,
\eqn\extrtwo{\phi (\sigma) =\left\{\matrix{-\lambda \sigma - {M \over
2\lambda}e^{-2\lambda \sigma} + \dots ,&\phi \rightarrow -\infty\cr
&\cr  -{1\over 4}log(\sigma -
\sigma_\infty),&\phi \rightarrow \infty}\right.}
and
\eqn\extrthree{e^{2\rho} = \left\{\matrix{1 -
{M\over \lambda}e^{-2\lambda\sigma} +
\dots,&\phi \rightarrow -\infty\cr &\cr
{const \over (\sigma - \sigma_\infty)^{5 \over 2}},&\phi \rightarrow
\infty}\right.}

This exhibits the features that as $\phi \rightarrow \infty$ the
curvature goes to zero and the distance to $\phi = \infty$ is infinite.

\newsec{Mean Field Equations}

At this point it would be natural to study the classical collapse of
matter coupled to our nonsingular versions of dilaton gravity.  However,
we have not been able to exactly integrate the equations of motion for
even the simplest choice of matter fields, massless $\bf f$ waves.
That being so, we decided to directly tackle the more ambitious problem
of incorporating our nonsingular black hole solutions
into a framework that allows us to follow the process of Hawking
evaporation to its endpoint.  It is difficult to do this in a way which
can be justified as a systematic approximation to the dynamics of
a four dimensional
dilaton black hole (as we have tried to imagine justifying the
modified lagrangians which we have studied up to this point).
The only available technique for studying Hawking evaporation is based
on the mean field equations of CGHS, which may be viewed as the leading
term in an expansion in the inverse number of massless matter fields in
the two dimensional effective lagrangian.  The
massless CGHS ${\bf f}$ fields will certainly be present in string theory, and
for large magnetic charge, there will be many of them.  However, in the
conventional large $N$ approximation, the string coupling is of order
$e^{2\phi} \sim {1\over N}$, and the modifications that we have made to
the CGHS lagrangian are formally of higher order in $1\over N$.
Furthermore, we would expect terms of the form $P(\phi ) (\nabla \bf
f)^2$ to be just as important as those we have included, and we do not
know how to evaluate the $\bf f$ functional integral in the presence of
such terms.  Finally, there are low energy
quantum corrections coming from the interaction of
the $\bf f$ fields with the two dimensional
remnant of the electromagnetic field \ref\lde{A. Peet, L. Susskind and
L. Thorlacius, ``Information loss and anomalous scattering'', Stanford
preprint, SU-ITP-92-16 (1992).}, which are also of the
same nominal order in $1\over N$ as our modifications of CGHS.

To make progress, we abandon a bit more of our pretense of realism, and
view the lagrangian
\eqn\efflag{{\cal L} = \sqrt{- g} [ {N\over\kappa}
 (D R + W) + (\nabla {\bf f})^2 ] }
as a ``two dimensional model field theory'' .  We also abandon the
identification of $e^{2\phi}$ with a coupling constant.  Indeed, in the
context of the effective lagrangian it is $D^\prime (\phi )$
rather than $e^{2\phi}$ which controls the magnitude of quantum
fluctuations of the graviton and dilaton.  We can now perform a
systematic large $N$ expansion of the theory presented in equation
\efflag .  We use the CGHS metric to regularize the $\bf f$ determinant
and obtain the mean field equations (in the conformal gauge):

\eqn\meanfielda{0 = D^{\prime\prime} \del_+\phi \del_-\phi +
D^\prime \del_+\del_-\phi + {1\over 4} e^{2(\rho - \phi)}W +
\kappa\del_+\del_-\rho}

\eqn\meanfieldb{0 = D^{\prime\prime} \del_+\phi \del_-\phi +
2 D^\prime \del_+\del_-\phi - D^\prime \del_+\del_-\rho - {W^\prime
\over 8} e^{2(\rho -\phi)} + {1\over 4} W e^{2(\rho -\phi)}}

\eqn\meanfieldc{\eqalign{T_{\pm\pm} = 0 = &(D^{\prime\prime} + 2 D^\prime)
\del_\pm\phi\del_\pm\phi + D^\prime \del^2_{\pm}\phi - 2 D^\prime
\del_\pm\phi \del_\pm\rho \cr + &  (\partial_{\pm}{\bf f})^2
- \kappa (\del_\pm\rho\del_\pm\rho -
\del^2_{\pm}\rho + t_{\pm})}}

\eqn\meanfieldd{\del_+\del_- {\bf f} = 0}
where $\rho$ is the CGHS Liouville field, ($\rho = \hat\rho + \phi$).

The kinetic term in these equations is nonsingular, so long as $2 \kappa
+ D^\prime < 0$ for all real values of $\phi$. For example, in the model
of section three, this requires $\kappa < 3 ({\gamma\over 2})^{2/3}$
This restriction ensures that
for all real values of $\phi$ the field space metric of the leading
order large $N$ effective lagrangian for graviton and dilaton,
is non-degenerate.
Thus, with this restriction, we do not expect to find the kind of
singularities that plague the large $N$ equations of CGHS \refs{\bddo ,\rst}

In \refs{\bddo ,\rst}  a signal of the possibility of a singularity for
generic initial conditions was found by fashioning the static equations
into the form,
\eqn\statsing{e^{-2\rho}\ddot\rho \sim {2(\dot\phi^2e^{-2\rho} -
\lambda^2) \over (1 - \kappa e^{2\phi})},}
which indicates the possibility of a curvature singularity at $\phi =
\half log(\kappa)$. It turned out that the static
solutions, quantum kinks of \refs{\bghs ,\hawk ,\sthandb} , do not have
a singularity at this
value of $\phi$, they bounce from $\phi$ just below the critical value
and have a weak coupling singularity. However, the value
$\phi = \half log(\kappa)$, is the position of the singularity in
solutions representing gravitational collapse. In our system the
corresponding equation is,
\eqn\statDW{e^{-2\rho}\ddot\rho \sim {(({W^\prime \over 2} + W)e^{-2\phi} -
D^{\prime\prime}\dot\phi^2 e^{-2\rho}) \over (D^\prime + 2 \kappa)},}
This expression exhibits no singular value of $\phi$ provided that $\kappa$
obeys the restriction, $2\kappa + D^\prime < 0$, for all $\phi$.

At the moment, we see no other recourse for understanding the solutions
of these equations than numerical work.  The equations describing
Hawking evaporation are hyperbolic partial differential equations of a
type notoriously resistant to accurate numerical analysis.  The
literature on numerical analysis of the related CGHS
equations \refs{\bghs ,\sthandb ,\hawk ,\hastew ,\dlowe} is full of
controversy rather than consensus.
We will attempt to numerically simulate the full nonlinear infall
problem, but there are several less ambitious things that can be done as
well.  We believe that it is reasonably easy to obtain very accurate
semi-analytical solutions to the static equations.  One can
then study small fluctuations around these solutions.  What we hope to
find is that the small fluctuation problem around static solutions with
ADM mass larger than the extremal value will contain negative modes
corresponding to the Hawking decay of these black holes, while small
fluctuations around the extremal solution are stable.  This would be
evidence that the extremal solution is a basin of attraction for
some region of initial conditions with ADM mass near the extremal value.
It would show that our candidate remnants are the endpoint of Hawking
evaporation for at least some region of black hole parameters.
We are presently engaged in setting up the numerical analysis of these
equations and hope to report on its results at an early date.

\newsec{Conclusions}

We have shown that a large class of Lagrangians of the form \lagone have
nonsingular black hole solutions with an infinite internal spacetime
hidden behind the horizon.  The essential criterion for this to occur
is that the kinetic term of the lagrangian be nonsingular, which
requires that the function $D^\prime (\phi )$ is nowhere vanishing.  We
must also impose certain restrictions on the behavior of the lagrangian
in the asymptotic regions $\phi\rightarrow\pm\infty$.  These ensure
correspondence with the CGHS model in the weak coupling regime,
boundedness of the curvature in all regions of spacetime, and the
infinite extent of the internal universe.

If the potential term $H$ has no zeroes, we find nonsingular solutions
for all values, both positive and negative, of the ADM mass, with no
dramatic change in behavior as the mass is varied.  We suspect that when
such models are coupled to quantum mechanical matter fields, they will
suffer (at least in the mean field approximation) from the problem of
unending Hawking radiation encountered in \negen .  If $H$ has some
number of simple zeroes, this problem may be avoided.  There is then an
{\it extremal} value of the ADM mass for which the horizons coincide,
leaving behind a nonsingular spacetime with zero Hawking temperature.
It is plausible that when these models
are coupled to matter, this extremal spacetime (or a slight quantum
deformation of it) will be a natural endpoint for Hawking evaporation.

Much work remains to be done to verify this conjecture.  We have not yet
even solved the classical equations for infalling matter in these
systems, but are instead engaged in a numerical study of the large
$N$ equations, which include both infall and backreaction. Preliminary
analysis suggests that these equations have static solutions
corresponding to nonsingular quantum deformations of the solutions
studied in this paper, and that they do not suffer from the
singularities discovered in \refs{\bddo ,\rst} .  We are quite concerned
however
about another sort of potential singularity.  As far as we can tell, all
of the spacetimes which arise in models in which $H$ has zeroes, have
Cauchy horizons.  It is widely believed that Cauchy horizons become
singular when subjected to small perturbations \ref\cauchy{S.
Chandrasekhar, and J. Hartle, ``On crossing the Cauchy
horizon of a Reissner-Nordstrom black hole,'' {\it Proc. Roy. Soc.
Lond.}, {\bf A384},(1982),301.} .  This suggests that
generic dynamical solutions of the mean field equations might have
singularities.  Thus, it is possible that our attempt to find a
nonsingular semiclassical description of black hole evaporation may
fail.

There is a bright side to this dismal conclusion.  Hawking's information
paradox was supposed to provide a first insight into the conceptual
problems of quantum gravity.  Its resolution by the agency of
cornucopions, or other large
remnants with small throats, is in some ways disappointingly
semiclassical.  It demonstrates once again\foot{The first demonstration
was the inflationary universe.} that a Hilbert space description of
quantum gravity must involve the notion of a changable number of states.
It may also lead to an argument that wormhole processes {\it must} be included
in any sensible theory of quantum gravity\foot{The cornucopion scenario
requires the description of the {\it asymptotic} states of geometry to
include disconnected spatial universes.  It is plausible that unitarity
and locality then require such disconnected universes to appear
in intermediate states as well.}.  However, the remnant scenario does
not seem to require us to understand the mind-boggling
prospects of large quantum fluctuations in the geometry of spacetime.
Nor does it require us to understand the generalization of geometry
provided by string theory.  Perhaps our (possible) failure to find an
adequate semiclassical description of black hole {\it singularities}
will force us to come to grips with these deep issues.
\vfill\eject
\listrefs

\end